\documentclass[aps,prd,twocolumn,showpacs,amsmath,amssymb,nofootinbib,eqsecnum, preprintnumbers]{revtex4-1}

\usepackage{bbm}
\usepackage{graphicx}
\usepackage{epstopdf}
\usepackage{color}

\newcommand{\eps}{\varepsilon}
\newcommand{\ph}{\varphi}

\DeclareMathOperator{\tr}{tr}

\begin{document}

\title{Geometry of Lax pairs: particle motion and Killing--Yano tensors}

\author{Marco Cariglia}
\email{marco@iceb.ufop.br}
\affiliation{Universidade Federal de Ouro Preto, ICEB, Departamento de F\'isica,   Campus Morro do Cruzeiro,
Ouro Preto 35400-000, MG,  Brasil}

\author{Valeri P. Frolov}
\email{vfrolov@ualberta.ca}
\affiliation{Theoretical Physics Institute,
University of Alberta,\\
Edmonton, Alberta, Canada T6G 2G7}

\author{Pavel Krtou\v{s}}
\email{Pavel.Krtous@utf.mff.cuni.cz}
\affiliation{Institute of Theoretical Physics,
Faculty of Mathematics and Physics, Charles University in Prague,
V~Hole\v{s}ovi\v{c}k\'ach 2, Prague, Czech Republic}

\author{David Kubiz\v n\'ak}
\email{dkubiznak@perimeterinstitute.ca}
\affiliation{Perimeter Institute, 31 Caroline St. N. Waterloo Ontario, N2L 2Y5, Canada}

\date{October 10, 2012}  

\begin{abstract}
A geometric formulation of the Lax pair equation on a curved 
manifold is studied using phase space formalism. The 
corresponding (covariantly conserved) Lax tensor is defined 
and the method of generation of constants of motion from it 
is discussed. It is shown that when the Hamilton equations 
of motion are used, the conservation of the Lax tensor 
translates directly to the well known Lax pair equation, with
one matrix identified with components of the Lax tensor and 
the other matrix constructed from the (metric) connection.
A generalization to Clifford objects is also discussed. 
Nontrivial examples of Lax tensors for geodesic and charged 
particle motion are found in spacetimes admitting hidden 
symmetry of Killing--Yano tensors.
\end{abstract}

\pacs{02.30.Ik, 02.40.Yy, 45.10.Na, 45.20.Jj, 04.50.Gh}
\preprint{pi-mathphys-299}

\maketitle

\section{Introduction}\label{sc:intro}
Since its discovery in 1968 \cite{Lax:1968}, the Lax pair formulation has played an invaluable role in studying integrability of various systems. Although first formulated for systems with infinite degrees of freedom, the formalism can be also used for, and provides an elegant description of, special finite dimensional systems with symmetries. Examples of such (completely integrable) systems admitting a Lax pair formulation include the Kepler problem, the Euler, Lagrange and Kowalevski tops, the Neumann model, or Toda lattice; we refer the reader to the monograph \cite{BabelonEtal:2003} and to references therein. In what follows we concentrate on finite dimensional systems. Namely, we shall discuss a geometrization of the Lax pair matrices for motion in curved spacetime.

A standard dynamical system is described on a phase space ${P}$ equipped with the symplectic 2-form $\Omega$ and with the corresponding Poisson brackets $\{,\}$. The dynamics is encoded in a Hamiltonian $H$ through the evolution equation for an arbitrary scalar observable ${F}$,
\begin{equation}\label{ur}
\dot{F}=\{F,H\}\, .
\end{equation}

The Lax pair method consists of finding two phase-space valued matrices $\mathsf{L}$ and $\mathsf{M}$, such that the equations of motion imply the {\em Lax pair equation}
\begin{equation}\label{Lax}
\dot{\mathsf{L}}=[\mathsf{L},\mathsf{M}]\;.
\end{equation}
The stronger formulation requires that Eq.~\eqref{Lax} implies the equations of motion, in which case the Lax pair formulation can be used as a starting point of the description of the dynamical system.

However, even without this latter stronger condition, the Lax pair matrices satisfying \eqref{Lax} play an important role in the study of integrability since they allow a simple construction of constants of motion. Indeed, the solution of \eqref{Lax} is of the form ${\mathsf{L}(t)=\mathsf{G}(t)\mathsf{L}(0)\mathsf{G}^{-1}(t)}$, where the evolution matrix ${\mathsf{G}(t)}$ is determined by the equation ${\dot{\mathsf{G}} = -\mathsf{M}\mathsf{G}}$. Therefore, if $I(\mathsf{L})$ is a function of $\mathsf{L}$ invariant under conjugation $\mathsf{L}\to \mathsf{G} \mathsf{L}\mathsf{G}^{-1}$, then $I(\mathsf{L}(t))$ is a constant of motion. All such invariants can be generated from the traces of various matrix-powers of $\mathsf{L}$:
\begin{equation}\label{traces}
\tr(\mathsf{L}^j)\;.
\end{equation}

The particular Lax pair may not yield all the constants of motion. However, in such a case it is often possible to upgrade the initial Lax pair so that the upgraded one already yields all the conserved observables of the dynamical system. Since the dimensionality of the Lax matrices is not fixed and the Lax pair equation is linear, two Lax pairs can be easily combined by their direct sum. Another useful method of producing a parametric class of Lax pairs is to introduce so-called spectral parameters, see, e.g., \cite{BabelonEtal:2003}.

Unfortunately, in general there is no constructive procedure  to find a Lax pair for the given problem or even to determine whether the Lax pair (in its stronger formulation) exists. Moreover, the solution is in no sense unique and even the dimensionality of the matrices may vary. However, when the Lax pair exists, it can be a very powerful tool for dealing with the conserved quantities.

In our paper, we focus on the construction of the Lax pair for motion on a curved manifold. In this case the phase space ${P}$ is given as a cotangent bundle of a configuration space ${M}$. There are thus two important features of such a theory which single it out among generic dynamical systems:
(i) It has preferred splitting of $2D$ phase coordinates into two sets---$D$ spacetime coordinates $x^a$ and $D$ momenta $p_a$.
(ii) The configuration space has usually an additional structure on it, the metric $g_{ab}$.

The prominent example of such dynamical systems is motion of particles and light in curved spacetime which plays an important role in General Relativity and its generalizations. We will concentrate mainly on this system. However, the dynamical systems with the cotangent bundle structure include also all non-relativistic systems which start with the Lagrangian description on the configuration space. The metric on the configuration space in such cases emerges from the kinetic part of the Hamiltonian.

Our aim is to show that for these dynamical systems it is possible to formulate a covariant analogue of the Lax equation. This covariant formulation allows us to employ the description of explicit and hidden symmetries of curved manifolds encoded in structures as Killing vectors, Killing tensors or Killing--Yano tensors.
In particular, for motion of particles in a curved spacetime we establish a relationship between conserved quantities, connected with the Lax pair, and integrals of motion connected with hidden symmetries, generated by Killing--Yano tensors.

To achieve this program, we first (in Sec.~\ref{sc:CovDerPhSp}) analyze geometrical structures on the phase space with cotangent bundle structure, especially those induced from a configuration-space covariant derivative. We define a covariant derivative acting on phase-space fields with configuration-space tensor indices.

With this geometrical background, in Sec.~\ref{sc:LaxPairs} we formulate the {\em covariant Lax equation}
\begin{equation}\label{CovLaxEqIntro}
    \frac{\nabla}{d t} L^{a}_{\,b} =0\;,
\end{equation}
for a phase-space dependent tensor field ${L^{a}_{\,b}(x,p)}$ and call this object a {\em Lax tensor}. Here, $\frac{\nabla}{dt}$ is the time derivative (the derivative along the Hamiltonian flow) defined in terms of the covariant derivative mentioned above.

By construction, any scalar invariant built covariantly from the Lax tensor (or from a set of Lax tensors) is preserved along  phase space trajectories and defines thus a constant of motion. Moreover, all scalar invariants encoded in one Lax tensor ${L^a{}_{\!b}}$ can be generated from the traces of powers of ${L^a{}_{\!b}}$,
\begin{equation}\label{traces2}
\tr(L^j)=\underbrace{L^a{}_{\!b} L^b{}_{\!c}\dots L^d{}_{\!a}}_{\mbox{\tiny{total of $j$ tensors}}}\;,
\end{equation}
which is an obvious geometric analogue of Eq.~\eqref{traces}.

There is a direct translation of the covariant Lax equation \eqref{CovLaxEqIntro} into the standard Lax pair equation \eqref{Lax}.
Namely, the components of Lax tensor ${L^{a}{}_b}$ form the Lax matrix
$\mathsf{L}$, while the other matrix $\mathsf{M}$ is given through the (Christoffel) connection symbols as follows:
\begin{equation}\label{LaxPairCoorIntro}
   \mathsf{L} = [L^{a}{}_{\!b}]\;,\qquad
   \mathsf{M} = \Bigl[\frac{\partial H}{\partial p_n}\Gamma_{\!nb}^a\Bigr]\;.
\end{equation}
An analogous geometric construction using Clifford matrices will be also introduced.
Similar ideas in special cases have been already studied in works \cite{Rosquist:1994,RosquistGoliath:1998, KarloviniRosquist:1999, BaleanuKarasu:1999, BaleanuBaskal:2000}.

The paper is organized as follows. In the next section we review the phase space formalism for the phase space built as a cotangent bundle of a configuration manifold. In particular, we define the covariant phase space derivative. This is further elaborated in the Appendix. In Sec.~\ref{sc:LaxPairs} we recapitulate the Lax pair formalism and derive its geometric covariant counterpart encoded in Eq.~\eqref{CovLaxEqIntro}.
Sec.~\ref{sc:TrivEx} illuminates the previous discussion by studying two `trivial' examples of Lax tensors for geodesic motion in generic spacetimes. Sec.~\ref{sc:KYCCKY} is devoted to highly non-trivial examples of Lax tensors for geodesic motion in special spacetimes admitting a hidden symmetry of Killing--Yano tensors.
The motion of a charged particle in weakly charged Kerr-NUT spacetimes is discussed in Sec.~\ref{sc:charged}. We conclude in Sec.~\ref{sc:summary}.

\section{Covariant derivative along a phase space trajectory}\label{sc:CovDerPhSp}

\subsection*{Phase space}
Particle motion in a curved spacetime\footnote{%
As we mentioned in Introduction, although we aim our discussion mainly to the relativistic context, the formalism introduced here does not depend on the signature of the metric and it can be used also in the case of  standard non-relativistic mechanics with a curved configuration space. We use small Latin letters for spacetime (configuration space) tensor indices and we drop indices when it does not cause confusion.}
${M}$ can be described in the language of Hamiltonian mechanics. The phase space ${P}$ is the cotangent bundle ${\textbf{T}^* M}$ equipped with the standard symplectic structure ${\Omega}$ and Poisson brackets. A phase space point can be written as ${[x,p]}$ with position ${x\in M}$ and momentum ${p\in\textbf{T}^*_x M}$. Any spacetime coordinates ${x^a}$ together with the corresponding components ${p_a}$ of the momentum ${p}$ form the canonical coordinates ${x^a,\,p_b}$ in which the symplectic structure and Poisson brackets read
\begin{gather}
   \Omega = d x^a \wedge d p_a \;,\label{SymplStrCoor}\\
   \{ F, \,G \}
     = \frac{\partial F}{\partial x^a} \frac{\partial G}{\partial p_a}
     - \frac{\partial F}{\partial p_a} \frac{\partial G}{\partial x^a}
     \;.\label{PoisBrCoor}
\end{gather}

Given a Hamiltonian ${H}$, the time derivative of any observable ${F}$ is
\begin{equation}\label{TimeDerObs}
  \dot F = \{F,\,H\} \;,
\end{equation}
which can be interpreted also as the derivative along the Hamiltonian flow ${X_{\!\textstyle H} = \Omega^{-1}\cdot dH}$\,,
\begin{equation}\label{HamFlow}
   \dot F = X_{\!\textstyle H} \cdot dF\;.
\end{equation}
More details on the conventions used for the symplectic structure and its inverse can be found in the Appendix.

\subsection*{Covariant phase space derivative}
The derivative \eqref{HamFlow} along ${X_{\!\textstyle H}}$ is defined only for \emph{scalar} phase-space observables. It turns out fruitful to generalize it to more general observables, namely to fields on phase space with spacetime indices.
Such fields appear naturally as a combination of spacetime tensors contracted with momenta, e.g.,
\begin{equation}\label{fieldexampl}
    A^{a\dots}_{b\dots}(x,p)
      = a^{a\dots}_{b\dots}(x)
      + b^{a\dots k}_{b\dots}(x)\, p_k
      + c^{a\dots kl}_{b\dots}(x)\, p_k p_l+\dots\;
\end{equation}
One can also consider fields of more complicated analytic form, e.g., ${A_a(x,p) = (g^{kl}(x)\, p_k p_l)^{-\frac12}\, p_a}$.

To define a derivative of tensor fields one needs an additional structure. For spacetime fields such a structure is a covariant derivative. Assuming the covariant derivative\footnote{%
This can be an arbitrary covariant derivative, not necessarily the metric one. For simplicity, in the body of the paper we assume vanishing torsion. Expressions with torsion can be found in the Appendix. In the case of geodesic motion we choose the metric covariant derivative.}
${\,\nabla}$ on the spacetime we lift this derivative to act on phase-space fields with spacetime indices. We call the resulting operation a {\em covariant phase space derivative}.

This derivative is defined along a general phase-space direction ${X\in\textbf{T}P}$. The direction can be represented using a configuration direction ${u\in\textbf{T}M}$ and a momentum direction ${f\in\textbf{T}^*M}$. The definition of this splitting requires the covariant derivative ${\nabla}$. Namely, if ${X}$ is tangent to a phase-space trajectory ${[x(t),p(t)]}$, its configuration part ${u}$ is tangent to the configuration trajectory ${x(t)}$ and the momentum part ${f}$ is the covariant derivative of ${p(t)}$ along ${x(t)}$. That is, we have  
\begin{equation}\label{finCoor}
    f_a = \frac{\nabla}{dt} p_a = \dot p_a - u^n \Gamma_{\!na}^k p_k \;.
\end{equation}
See Eq.~\eqref{apx:fpdotrel} and Fig.~\ref{fig:phspspl} in Appendix for further discussion.

The covariant derivative ${\nabla_{\!\textstyle X}}$ along the direction ${X=[u,f]}$ acting on phase-space fields with spacetime indices is defined by the following rules:
\begin{enumerate}
\item[(i)]
  For a field depending only on the spacetime position, ${A^{a\dots}_{b\dots}(x,p)=\alpha^{a\dots}_{b\dots}(x)}$, the derivative reduces to the standard covariant derivative along ${u}$:
  \begin{equation*}
    \nabla_{\!\textstyle X}\, A^{a\dots}_{b\dots}
       = u^n \nabla_{\!n}\, \alpha^{a\dots}_{b \dots}\;.
  \end{equation*}
\item[(ii)]
  For the momentum field ${p_a}$, the derivative gives the momentum part of ${X}$:
  \begin{equation*}
    \nabla_{\!\textstyle X}\, p_a = f_a\;.
  \end{equation*}
\item[(iii)]
  The derivative ${\nabla_{\!\textstyle X}}$ satisfies all standard rules for the derivative (i.e., linearity, the Leibniz product rule, and the chain rule).
\end{enumerate}

These rules reflect the splitting of the phase-space direction into configuration and momentum parts. This can be also encoded using the partial derivatives\footnote{%
In short, the partial derivative ${\frac{\partial}{\partial p}}$ is the derivative in a momentum direction with ${x}$ fixed and ${\frac{\nabla}{\partial x}}$ is the derivative in a configuration direction with ${p}$ parallel-transported. For more details, see \eqref{apx:momder} and \eqref{apx:posder}.}
${\frac{\nabla}{\partial x}}$ and ${\frac{\partial}{\partial p}}$ introduced in Appendix:
\begin{equation}\label{DerSpl}
  \nabla_{\!\textstyle X} A^{a\dots}_{b\dots}
    = u^n \frac{\nabla_{\!n} }{\partial x} A^{a\dots}_{b\dots}
    + f_n \frac{\partial}{\partial p_n} A^{a\dots}_{b\dots}\;.
\end{equation}
For a phase-space trajectory with tangent field ${X}$ we naturally write
\begin{equation}\label{nabladt}
    \frac{\nabla}{d t} A^{a\dots}_{b\dots}
       \equiv \nabla_{\!\textstyle X} A^{a\dots}_{b\dots}\;,
\end{equation}
and call ${\frac{\nabla}{d t}}$ the {\em covariant derivative along the phase space trajectory} or just the {\em covariant time derivative}.
The configuration and momentum parts of the Hamiltonian flow ${X_{\!\textstyle H}}$ are
\begin{equation}\label{hamflowsplit}
   u^a = \frac{\partial H}{\partial p_a}\;,\qquad
   f_a = -\frac{\nabla_{\!a} H}{\partial x}\;,
\end{equation}
cf.\ Eq.~\eqref{apx:hamflowsplit}, which gives
\begin{equation}\label{DerSplHamF}
  \frac{\nabla}{dt} A^{a\dots}_{b\dots}
    = \frac{\partial H}{\partial p_n} \, \frac{\nabla_{\!n} }{\partial x} A^{a\dots}_{b\dots}
    - \frac{\nabla_{\!n} H}{\partial x} \frac{\partial}{\partial p_n} A^{a\dots}_{b\dots}\;.
\end{equation}
This is a natural generalization of Eq.~\eqref{TimeDerObs} to the case of  tensor-valued observables.

The introduced phase-space covariant derivative can be expressed in coordinates,
\begin{equation}\label{CovDerPhSpCoor}
  \nabla_{\!\textstyle X} A^{a\dots}_{b\dots}
     = \dot A^{a\dots}_{b\dots}
     + u^n \Gamma_{\!nk}^a A^{k\dots}_{b\dots} + \dots
     - u^n \Gamma_{\!nb}^k A^{a\dots}_{k\dots} - \dots\,,
\end{equation}
where ${\dot A^{a\dots}_{b\dots}}$ is the derivative of components of ${A}$ along ${X}$ direction,
\begin{equation}\label{CoorDerPhSp}
  \dot A^{a\dots}_{b\dots}
    = u^n \frac{\partial A^{a\dots}_{b\dots}}{\partial x^n}
    + \dot p_n \frac{\partial A^{a\dots}_{b\dots}}{\partial p_n} \;.
\end{equation}
Note that the coordinate time derivative ${{\dot p}_a}$ and the covariant derivative ${f_a}$ are related by Eq.~\eqref{finCoor}.

For more details on the derivatives ${\nabla_{\!\textstyle X}}$, ${\frac{\nabla}{\partial x}}$, ${\frac{\partial}{\partial p}}$ and the corresponding coordinate expressions we refer the reader to the Appendix.

\subsection*{Derivative of Clifford fields}
The spacetime metric derivative can also be lifted to a derivative acting on phase-space fields with Dirac spinor indices. Namely, we are interested in Clifford objects, i.e., operators acting on Dirac spinors. These are generated by the abstract gamma matrices ${\gamma^a}$, obeying\footnote{%
In expressions with Clifford objects and Dirac spinors the Clifford multiplication is assumed. In components, it reduces to the standard matrix multiplication.}
\begin{equation}\label{Clifford}
    \gamma^a \gamma^b + \gamma^b \gamma^a = 2\,g^{ab} \mathbbm{1} \;.
\end{equation}
Thanks to this rule, any Clifford object ${\slash\!\!\!\omega}$ can be represented by an inhomogeneous antisymmetric form ${\omega=\sum_r {}^r\!\omega}$\,,
\begin{equation}\label{ClExt}
    \slash\!\!\!\omega
      = \sum_r \frac1{r!}\, {}^{r}\!\omega_{a_1\dots a_r} \gamma^{a_1\dots a_r}\;.
\end{equation}
Here, ${}^r\!\omega$ are homogeneous rank-$r$ antisymmetric forms and
\begin{equation}\label{gammamult}
    \gamma^{a_1\dots a_r} = \gamma^{[a_1}\cdots\gamma^{a_r]}\;.
\end{equation}

The covariant derivative on the Dirac bundle is induced from the spacetime metric derivative by the condition\footnote{%
To define the covariant derivative on Dirac spinors uniquely, the condition \eqref{Dergamma0} must be supplemented by some further conditions reflecting irreducibility and reality properties of ${\gamma^a}$. However, thanks to Eq.~\eqref{Dergamma0} and rule \eqref{Clifford}, only the covariant derivative which annihilates the metric can be lifted to the Dirac bundle. In this context we always assume vanishing torsion, so the derivative on tangent bundle must be the metric derivative.}
\begin{equation}\label{Dergamma0}
    \nabla_{\!n} \gamma^a = 0\;.
\end{equation}
Clearly, we can lift this derivative to act on the Clifford valued fields on the phase space in a similar way as we did for spacetime-tensor valued fields. Namely, for ${\Lambda(x,p)=\lambda_{a_1\dots a_r}(x,p)\;\gamma^{a_1\dots a_r}(x)}$
we simply get
\begin{equation}\label{ClCovDerPhSp}
    \nabla_{\!\textstyle X}\Lambda
      = \bigl(\nabla_{\!\textstyle X} \lambda_{a_1\dots a_r}\bigr)
        \, \gamma^{a_1\dots a_r}\;.
\end{equation}

To write down this covariant derivative in components, in addition to coordinates ${x^a}$, one has to introduce an orthonormal frame ${e_{\hat n}\in\mathbf{T}M}$ and the spinor frame ${E_\Upsilon}$ in such a way that the components\footnote{%
Components with respect to the frame ${e_{\hat n}}$ will be denoted with hatted indices. We will mostly skip the spinor indices (capital Greek letters), i.e., instead of ${\gamma^{\hat a}{}^\Phi{}_\Psi}$ we write just ${\gamma^{\hat a}}$. The matrix multiplication between Clifford objects and spinors is assumed.}
${\gamma^{\hat a}}$ of the gamma matrices are constants. The covariant derivative of the Dirac spinor ${\Phi}$ expressed in the spinor frame then reads
\begin{equation}\label{CovDerDirac}
    \nabla_{\!a} \Phi = \Phi_{,a} + \Sigma_a{} \Phi\;,
\end{equation}
with the connection coefficients ${\Sigma_a}$ uniquely determined in terms of the Ricci coefficients
${\hat\Gamma^{\hat m}_{\!a\hat n}=(\nabla_{\!a} e_{\hat n}^k)\, e^{\hat m}_k}$
by the standard relation
\begin{equation}\label{SigmaDirac}
    \Sigma_a = \frac14\, \hat\Gamma^{\hat m}_{\!a\hat n} \;
               \gamma{}_{\hat m}{}^{\hat n}\;.
\end{equation}

With these definitions the covariant derivative of the Clifford field ${\Lambda(x,p)}$ on the phase space is
\begin{equation}\label{ClCovDerCoor}
    \nabla_{\!\textstyle X} \Lambda
      = \dot \Lambda
      + [ \, u^n \Sigma_{n} ,\, \Lambda\,]\;,
\end{equation}
where ${\dot\Lambda}$ is just an ordinary derivative along ${X}$ of components ${\Lambda^\Phi{}_\Psi}$.

\section{Covariant Lax equation}\label{sc:LaxPairs}

\subsection*{Lax pair}
The Lax pair provides a useful tool for generating conserved quantities. The phase-space valued matrices ${\mathsf{L}}$ and ${\mathsf{M}}$ form the Lax pair if they satisfy the Lax pair equation~\cite{Lax:1968}
\begin{equation}\label{LaxPairEq}
    \dot{\mathsf{L}} = [\mathsf{L},\,\mathsf{M}]\;.
\end{equation}
Here, the dot is understood as the ordinary time derivative of each component of the matrix ${\mathsf{L}}$. It follows that any scalar invariant formed from the matrix ${\mathsf{L}}$ is a conserved quantity, cf.\ Eq.~\eqref{traces}.

It is customary to require that the Lax pair satisfies additional properties, especially, that (i) Eq.~\eqref{LaxPairEq} is equivalent to the equations of motion, and that (ii) the invariants of ${\mathsf{L}}$ generate the maximal number of conserved quantities of the system. However, we will study Lax pairs without requiring these additional conditions. This is justified by realizing that the Lax-pair equation \eqref{LaxPairEq} is linear in both ${\mathsf{L}}$ and ${\mathsf{M}}$. One can thus obtain a `more sophisticated' Lax pair as a direct sum of smaller matrices, each of which satisfy Eq.~\eqref{LaxPairEq}, and impose additional conditions only at the end, on the resulting pair.

\subsection*{Lax tensor}
We shall now formulate an alternative covariant description of the Lax equation and clarify its relation to the standard Lax pair formulation.
Using the above definition of the covariant phase space derivative, we define the {\em Lax tensor} to be a covariantly conserved tensor field ${L^{a}_{\,b}(x,p)}$, obeying the {\em covariant Lax tensor equation}
\begin{equation}\label{CovLaxEq}
    \frac{\nabla}{d t} L^{a}_{\,b} =0\;.
\end{equation}
Obviously, any scalar covariantly constructed\footnote{\label{ftn:CovDer}%
The covariant derivative employed in the Lax equation \eqref{CovLaxEq} can be arbitrary. By covariant construction then we mean any operation which commutes with this derivative. Typical covariant operations are traces, contracted multiplication, or the determinant. If a tensor covariantly constant with respect to the chosen derivative is available, it can be used to construct the conserved scalar. Therefore, we typically choose the metric covariant derivative since then the metric ``is available''. However, one could use a different derivative, for example, if the Hamiltonian is ${H=\frac12p_ap_bk^{ab}}$ with the ``inverse mass'' tensor ${k^{ab}}$ different from the metric ${g^{ab}}$.}
from the Lax tensor or a set of Lax tensors is a constant of motion. In particular, this is true for invariants constructed as traces \eqref{traces2} of various powers of $L^a{}_b$.
Similar to the Lax pair, the Lax tensor hence generates constants of motion.

Moreover, each Lax tensor defines a Lax pair. Indeed, in
 components, while using \eqref{CovDerPhSpCoor}, Eq.~\eqref{CovLaxEq} implies
\begin{equation}\label{CovConsCoor}
    \dot L^{a}_{\,b}
    = L^{a}_{\,k}\, u^n\Gamma_{\!nb}^k
    - u^n\Gamma_{\!nk}^a \, L^{k}_{\,b}
    \;.
\end{equation}
This equation is in form already very close to \eqref{LaxPairEq}. However, the matrices forming the Lax pair must be defined as functions on phase space. Therefore we have to eliminate the velocity ${u}$ using the first of the Hamilton equations \eqref{hamflowsplit}. The correspoding Lax pair matrices thus are
\begin{equation}\label{LaxPairCoor}
   \mathsf{L} = [L^{a}_{\,b}]\;,\qquad
   \mathsf{M} = \biggl[\frac{\partial H}{\partial p_n}\,\Gamma_{\!nb}^a\biggr]\;.
\end{equation}
Hence, equation \eqref{CovLaxEq} can be understood as a `covariant generalization' of the Lax pair equation \eqref{LaxPairEq};
its coordinate form gives the Lax pair in the ordinary sense.

For a motion in curved space governed by a given Hamiltonian, the covariant Lax tensor satisfying \eqref{CovLaxEq} and the ordinary Lax pair matrices \eqref{LaxPairCoor} carry the same information. The Lax pair matrices are, however, coordinate dependent. Under a change of coordinates the matrix ${\mathsf{L}}$ transforms just by a trivial conjugation. However, the matrix ${\mathsf{M}}$ changes in a more complicated manner since the connection coefficients are involved.


\subsection*{Clifford Lax tensor}
Any covariantly conserved\footnote{%
In the context of Clifford fields we always assume the metric covariant derivative since the abstract gamma matrices are then covariantly constant \eqref{Dergamma0}, cf.\ also footnote \ref{ftn:CovDer}.}
antisymmetric form ${\lambda_{ab\dots}}$ induces the Clifford field ${\Lambda = \lambda_{ab\dots} \gamma^{ab\dots}}$ satisfying
\begin{equation}\label{CovConsCl}
    \frac{\nabla}{d t} \Lambda = 0\;.
\end{equation}
We call such $\Lambda$ a {\em Clifford Lax tensor}.
In components, using Eq.~\eqref{ClCovDerCoor}, Eq.~\eqref{CovConsCl} implies
\begin{equation}\label{CovConsClCoor}
  \dot \Lambda
      = [ \,\Lambda,\, u^n \Sigma_n\,]\;.
\end{equation}
In a way similar to previous subsection we find that matrices
\begin{equation}\label{LaxPairCl}
  \mathsf{L} = \bigl[\Lambda\bigr]
    = \bigl[\lambda_{\hat a\hat b\dots}\gamma^{\hat a\hat b\dots}\bigr]
  \;,\quad
  \mathsf{M}
    = \biggr[\frac{\partial H}{\partial p_n}\,\Sigma_n\biggl]\;,
\end{equation}
form a Lax pair satisfying \eqref{LaxPairEq}.

In the next two sections we shall give a number of examples of (Clifford) Lax tensors for geodesic motion in curved spacetime.
Whereas the following section concentrates on `trivial' examples in generic spacetimes (no enhanced symmetry is assumed), in Sec.~\ref{sc:KYCCKY}
we discuss  Lax tensors in special spacetimes admitting hidden symmetries. The motion of a charged particle in weakly charged Kerr-NUT spacetimes
is discussed in Sec.~\ref{sc:charged}.

\section{Lax tensors and geodesic motion: two trivial examples}\label{sc:TrivEx}

\subsection*{Geodesic motion}
A geodesic motion with respect to the spacetime metric ${g_{ab}}$ is governed by the Hamiltonian
\begin{equation}\label{geodHamilt}
    H = \frac1{2m}\, p_a g^{ab} p_b\;.
\end{equation}
The equations of motion \eqref{TimeDerObs} for canonical coordinates ${x^a,\, p_a}$ are
\begin{equation}\label{geodincoor}
    \dot x^a = m^{-1} g^{an} p_n\;\qquad
    \dot p_a = - \frac1{2m} \frac{\partial g^{kl}}{\partial x^a}\, p_k p_l\;.
\end{equation}
These equations are equivalent to the geodesic equation ${u^n\nabla_{\!n} u^a=0}$  with the velocity ${u^a=\dot x^a}$ and ${\nabla}$ being the metric covariant derivative. They are also equivalent to the covariant equations on the phase space:
\begin{equation}\label{GeodPhSp}
    p_a = m\, g_{an} u^n\;,\qquad
    \frac{\nabla}{dt} p_a = 0\;.
\end{equation}

\subsection*{Lax tensor implying geodesic motion: Example I}

As a first example, let us study the simple tensor
\begin{equation}\label{Lpp}
    L^a{}_{\!b} = g^{an} p_n p_b\;.
\end{equation}
Thanks to Eq.~\eqref{GeodPhSp}, it is covariantly conserved and hence defines a Lax tensor.

In fact, we shall prove now that the existence of this Lax tensor is equivalent to the geodesic motion and hence it can be used as a starting point of the dynamics. In order to do that, we investigate the following problem: Assuming that the covariant Lax equation \eqref{CovLaxEq} along an unknown Hamiltonian flow ${X_{\!\textstyle H}}$ is satisfied for \eqref{Lpp}, does it imply that the Hamiltonian ${H}$ must generate geodesic motion?

Substituting \eqref{Lpp} into \eqref{CovLaxEq}, we obtain
\begin{equation}
    \frac{\nabla}{d t} L^a_{\,b} = f^a p_b + p^a f_b = 0\;,
\end{equation}
which for generic momentum ${p}$ implies ${f=0}$. However, for the Hamiltonian flow, ${f}$ is given by \eqref{hamflowsplit} and we have
\begin{equation}
    f = -\frac{\nabla H}{\partial x} = 0\;.
\end{equation}
The Hamiltonian must thus be constructed only from momenta and covariantly constant spacetime tensors.
In a generic curved spacetime the only covariantly constant spacetime tensors are constructed from the metric. Therefore, the Hamiltonian must be of the form
\begin{equation}\label{HamiltL=pp}
    H = \frac12 h(p^2)\;,\qquad p^2 = p_a p_b g^{ab} \;,
\end{equation}
with ${h}$ being an arbitrary function. It implies the velocity ${u^a=h'(p^2) g^{an} p_n}$. Since ${p^2}$ is conserved, we reproduced the geodesic equations of motion.

The significance of the Lax tensor \eqref{Lpp} lies thus in the fact that it
implies geodesic motion. On other hand, the only conserved quantities which can be obtained from this Lax tensor are functions of ${p^2}$ (which is, of course, the only conserved quantity for generic geodesic motion).

Using \eqref{LaxPairCoor}, the Lax tensor \eqref{Lpp} defines the following Lax pair matrices:
\begin{equation}\label{LppLaxPair}
   L^a{}_{\!b} = g^{an} p_n p_b\;,\qquad
   M^a{}_{\!b} = \frac1m\, p_k\, g^{kl}\, \Gamma^{a}_{\!lb}\;.
\end{equation}
We could ask the same question about equations of motion when starting with the ordinary Lax pair \eqref{LppLaxPair} satisfying \eqref{LaxPairEq}. In this case the situation is slightly different. Using the explicit formula for the Christoffel coefficients ${\Gamma^{a}_{\!lb}}$ one can show that
\begin{equation}\label{LM=PBr}
    \bigl[ L,\, M \bigr]^a{}_{\!b}
    = \Bigl\{ L^{a}{}_{\!b},\, \frac1{2m} p^2 \Bigr\}\;,
\end{equation}
where the right-hand side is the Poisson bracket of the components of the Lax tensor ${L}$. Since ${\dot L^a{}_{\!b} = \{L^a{}_{\!b},\,H\}}$, the Lax pair equation \eqref{LaxPairEq} implies the following condition on the Hamiltonian:
\begin{equation}\label{LppHamCond}
    \Bigl\{ g^{an} p_{n} p_{b},\, H - \frac1{2m} p^2 \Bigr\} = 0\;.
\end{equation}
Obviously, the geodesic Hamiltonian ${H=\frac1{2m}p^2}$ solves this condition. However, it is not clear if this is a unique solution.


\subsection*{Lax tensor implying geodesic motion: Example II}

Allowing for Clifford-valued fields, it is possible to write an even simpler Lax tensor for the geodesic motion:\footnote{%
This tensor is in some sense `Dirac's square root' of the Lax tensor \eqref{Lpp}; it can be obtained from the WKB approximation of the Dirac equation.}
\begin{equation}\label{Lpg}
    \Lambda = p_a \gamma^a\;.
\end{equation}
Similar to the Example I, 
this Lax tensor exists in a generic spacetime, it generates only conserved quantities which are functions of $p^2$, and the corresponding covariant Lax equation is equivalent to the geodesic equations of motion.

Using \eqref{LaxPairCl}, the corresponding Lax pair matrices are
\begin{equation}\label{LpgLaxPair}
\begin{gathered}
   \mathsf{L} = \bigl[ \Lambda \bigr]
     =\bigl[ p_{\hat a}\gamma^{\hat a} \bigr]\;,\\
   \mathsf{M} = \Bigl[\frac1m\, p^n\, \Sigma_{n}\Bigr]
     = \Bigl[\frac1{4m}\, p^n\, \hat\Gamma_{\!n\hat k\hat l}\gamma^{\hat k\hat l}\Bigr]\;.
\end{gathered}
\end{equation}
Similarly to Example I, it is not clear to us whether these Lax pair matrices imply geodesic motion. Namely,
the commutator in the Lax equation \eqref{LaxPairEq} can be simplified using properties of the gamma matrices
\begin{equation}\label{LpgLaxPairCoor}
   \dot p_{\hat a} \gamma^{\hat a}
     = \frac1m p^m p_{\hat n} \hat\Gamma^{\hat n}_{\!m\hat a} \gamma^{\hat a}\;.
\end{equation}
We can thus eliminate the gamma matrices from both sides. Substituting the expression for the Ricci coefficients in terms of the derivatives of components of ${e_{\hat n}}$, it is possible to show that
\begin{equation}\label{phatM=PBr}
    \frac1m p^m p_{\hat n} \hat\Gamma^{\hat n}_{\!m\hat a}
      = \Bigl\{ p_{\hat a}, \frac1{2m}p^2\Bigr\}\;.
\end{equation}
Together with ${\dot p_{\hat a} = \{p_{\hat a}, H\}}$ it leads to an equation analogous to Eq.~\eqref{LppHamCond},
\begin{equation}\label{LpgHamCond}
    \Bigl\{ p_{\hat a},\, H - \frac1{2m} p^2 \Bigr\} = 0\;.
\end{equation}
Beware however, that the frame component ${p_{\hat a}}$ of the momentum is not the canonical coordinate conjugated with ${x^a}$ (indeed, ${p_{\hat a} = p_n e^n_{\hat a}(x)}$), so \eqref{LpgHamCond} does not imply that ${H-\frac1{2m}p^2}$ is independent of the coordinates. Similarly to the previous case, the geodesic Hamiltonian solves this condition but it is an open question whether this solution is unique.

\section{Lax tensors and hidden symmetries}\label{sc:KYCCKY}
Until now we have investigated rather trivial examples of Lax tensors, constructed from momenta and  metric-related quantities, 
and hence trivially conserved along the geodesic motion. Such objects are present in any generic spacetime.
Now we shall turn to examples of Lax tensors present in  spacetimes with enhanced symmetries, in particular admitting hidden symmetries of Killing and Killing--Yano tensors.

\subsection*{Lax tensor from conserved quantity}
Let us assume that, provided a given enhanced symmetry of the spacetime, an additional conserved quantity ${E}$ for geodesic motion is known.
This, for example, incorporates the case of $E$ generated from a Killing vector $\xi$, ${E=\xi^ap_a}$, or $E$ generated from a Killing tensor $k$,
${E=k^{ab\dots}\,p_a p_b\dots}$. Of course, ${L=E}$ is a trivial one-dimensional Lax tensor, which can be helpful, for example, if one constructs a larger Lax matrix as a sum of independent segments.
One can also simply upgrade the Lax tensors \eqref{Lpp} or \eqref{Lpg} as
\begin{equation}\label{LppE}
    L_E{}^a{}_{\!b} = E\, p^a p_b\,,\quad \Lambda_E=E p_a \gamma^a\,.
\end{equation}
The corresponding Lax pair matrices follow from \eqref{LaxPairCoor} or \eqref{LaxPairCl}.
If more than one such constants are known, we may employ the method of spectral parameters \cite{BabelonEtal:2003} to combine the corresponding Lax tensors.

%

%

\subsection*{Two kinds of Killing--Yano tensors}

The {\em Killing--Yano} (KY) tensor ${\ph_{a_1\dots a_r}}$ is \cite{Yano:1952} an antisymmetric form on the spacetime, the covariant derivative of which is determined by its antisymmetric part, i.e., by its exterior derivative ${d\ph}$:
\begin{equation}\label{KYdef}
  \nabla_{\!a} \ph_{a_1\dots a_r} = \nabla_{\![a} \ph_{a_1\dots a_r]}\;.
\end{equation}
On the  other hand, the {\em closed conformal Killing--Yano} (CCKY) tensor ${h_{a_1\dots a_r}}$ is an antisymmetric form on the spacetime, the covariant derivative of which is determined by its divergence ${\xi_{a_2\dots a_r}}$:
\begin{equation}\label{CCKYdef}
\begin{gathered}
  \nabla_{\!a} h_{a_1a_2\dots a_r} = r\, g_{a[a_1} \xi_{a_2\dots a_r]}\;,\\
  \xi_{a_2\dots a_r} = {\textstyle\frac{1}{D-r+1}} \nabla_{\!n} h^n{}_{a_2\dots a_r}\;.
\end{gathered}
\end{equation}
KY and CCKY tensors are related to each other through the Hodge duality: the Hodge dual of a KY form is a CCKY form and vice versa.

Both KY and CCKY tensors encode the so called hidden symmetries: they exist only for special metrics and provide a rather rich structure to the geometry, see, e.g.,
\cite{Semmelmann:2002, Papadopoulos:2008, FrolovKubiznak:2008, YasuiHouri:2011, Santillan:2011, Visinescu:2012, Cariglia:2012}
and references therein.

\subsection*{Covariantly conserved tensors}

As one consequence of the above definitions, both KY and CCKY tensors define a tensorial quantity linear in momentum which is conserved along trajectories of the geodesic motion. Namely, for a KY tensor ${\ph}$ and a CCKY tensor ${h}$ the quantities\footnote{%
The dot ``${\cdot}$'' denotes the contraction, ${(\ph\cdot p)_{a_1\dots a_r} = \ph_{a_1\dots a_rn}p^n}$. Since we assume automatic rising of indices, it is essentially the scalar product.}
\begin{equation}\label{2forms}
\kappa={\ph\cdot p}\,,\quad \mu={h\wedge p}\,,
\end{equation}
are conserved along geodesics,
\begin{equation}\label{fphpCovCons}
    \frac{\nabla}{d t} \kappa= 0\;,\quad
    \frac{\nabla}{d t} \mu = 0\;.
\end{equation}
The velocity is given by ${u=\frac1m p}$ and the momentum ${p}$ is conserved for the geodesic motion. Thus, thanks to \eqref{KYdef}, the derivative of $\kappa$ gives ${\frac1m p^k(\nabla_{\![k}\ph_{a_1\dots a_{r{-}1}l]})p^l=0}$. Similarly, Eq.~\eqref{CCKYdef} implies ${\frac\nabla{dt}h = \frac1m p\wedge h}$ which vanishes when wedged with another ${p}$. Of course, both equations in \eqref{fphpCovCons} are equivalent through the Hodge duality: ${\kappa= *\mu}$ for ${\ph=*h}$.

It is interesting to observe that the same information encoded in the forms $\kappa$ and $\mu$ is also encoded in the forms
\begin{equation}\label{PhiFdef}
\Phi = \kappa\wedge p\;,\quad F=\mu\cdot p\;,
\end{equation}
respectively. Indeed, employing the identity
\begin{equation}\label{dw+wd=sq}
    (\alpha\cdot p)\wedge p + (\alpha \wedge p)\cdot p = p^2\,\alpha\;,
\end{equation}
which holds for any antisymmetric form ${\alpha}$, one can reconstruct $\kappa$ and $\mu$ from the quantities above
\begin{equation}\label{reconstruct}
\begin{gathered}
\kappa= \frac{1}{p^{2}}\,\Phi\cdot p\;,\quad
\mu= \frac{1}{p^{2}}\,F\wedge p\;.
\end{gathered}
\end{equation}
Since the momentum is conserved for the geodesic motion, we thus obtained conserved quantities quadratic in momenta
\begin{equation}\label{fpphppCovCons}
    \frac{\nabla}{d t} \Phi = 0\;,\quad
    \frac{\nabla}{d t} F = 0\;,
\end{equation}
which, however, carry the same information as those defined in \eqref{2forms}. Note also the Hodge duality ${\Phi=*F}$ for ${\ph=*h}$.

Finally, the form ${F}$ can be rewritten using \eqref{dw+wd=sq} as
\begin{equation}\label{hwedgepdotp}
   F = (h\wedge p)\cdot p = p^2 h - (h \cdot p) \wedge  p\;,
\end{equation}
which in indices reads
\begin{equation}\label{FisPh}
\begin{split}
    &F_{a_1a_2\dots a_r} = \bigl((h\wedge p)\cdot p\,\bigr)_{a_1a_2\dots a_r}
    \\&\quad= p^2 h_{a_1a_2\dots a_r}
       -  h_{a_1 a_2\dots a_{r-1}n} p^n p_{a_r}
      \\ &  \mspace{50mu} - (-1)^{r-1} h_{a_2 a_3\dots a_{r}n} p^n p_{a_1}
       - \dots
    \\&\mspace{50mu}
       - (-1)^{r-1} h_{a_r a_1\dots a_{r-2}n} p^n p_{a_{r-1}}
    \\&\quad= p^2\; h_{n_1n_2\dots n_r}\;
       P^{n_1}{}_{\!\!a_1} P^{n_2}{}_{\!\!a_2} \dots P^{n_r}{}_{\!\!a_r}\;,
\end{split}\raisetag{3ex}
\end{equation}
where we have introduced the {\em projector}
\begin{equation}\label{Pdef}
    P^a{}_{\!b} = \delta^a{}_{\!b} - p^{-2} p^a p_b\;.
\end{equation}
The form ${F}$ is thus (up to the prefactor ${p^2}$) the projection of the CCKY tensor ${h}$ onto a subspace orthogonal to the momentum ${p}$.

\subsection*{Killing--Yano and Lax tensors}

All the conserved tensorial quantities $\kappa$, $\mu$, ${\Phi}$, and ${F}$ constructed from KY and CCKY tensors can be converted into Lax tensors using gamma matrices. For example, taking $\kappa$ and $\mu$, the corresponding Clifford objects read\footnote{Similar to Example II in the previous section, these Lax tensors can be understood as
arising from the WKB approximation to the Dirac symmetry operators $K_\ph$ and $M_h$ studied in \cite{CarigliaEtal:2011a, CarigliaEtal:2011b}.}
\begin{align}\label{FLaxTens}
   \Lambda_\ph &= \gamma^{a_1\dots a_{r-1}}(p\cdot \ph)_{a_1\dots a_{r-1}}\;,\\
   \Lambda_h &= \gamma^{a_1\dots a_{r+1}}(h\wedge p)_{a_1\dots a_{r+1}}\;.
\end{align}
They are covariantly conserved \eqref{CovConsCl} and the corresponding Lax pair of matrices are given by \eqref{LaxPairCl}. As mentioned above, the invariants of ${\Lambda_\ph}$ span the same functional space as invariants of ${\Phi}$, in other words, the conserved scalar observables generated from the Lax tensor ${\Lambda}_\ph$ are the same as those generated directly from $\Phi$, and similarly for ${\Lambda}_h$ and $F$.

Among all conserved quantities constructed from KY or CCKY tensors there are special cases which do not need to use Clifford objects and gamma matrices. Such a situation occurs if the conserved tensor under consideration is of  rank 2. This includes:

(a) The CCKY tensor ${h_a}$ of rank 1, which is in fact a closed conformal Killing vector
\begin{equation}\label{CCKv}
    \nabla_{\!a} h_{b} = \xi\, g_{ab}\;,\qquad
    \xi = \frac1D\,\nabla_{\!n}h^n\;.
\end{equation}
The tensor $\mu$ is now of rank 2, and hence generates directly the Lax tensor
\begin{equation}\label{Lhp}
    L^a{}_{\!b} = h^a p_b - p^a h_b\;.
\end{equation}
The Lax pair matrices are given by
\begin{equation}\label{LgeodLaxPairCoor}
    \mathsf{L} = \bigl[ L^a{}_{\!b} \bigr]\;,\qquad
    \mathsf{M} = \biggl[ \frac1m p^n\Gamma^{a}_{\!nb}\biggr]\;,
\end{equation}
cf.\ Eq.~\eqref{LaxPairCoor}. The only independent constant of motion which can be obtained from this Lax tensor is ${p^2\, h\cdot P \cdot h}={p^2 h^2 - (p\cdot h)^2}$.

(b) The CCKY tensor ${h_{ab}}$ of rank 2 generates the second-rank conserved quantity ${F_{ab}}$ given by \eqref{PhiFdef} or \eqref{FisPh}. The Lax tensor thus reads
\begin{equation}\label{LF}
    L^a{}_{\!b} = F^a{}_{\!b} = p^2 h^a{}_{\!b} - p^a p^n h_{nb} - h^{an}p_n p_b\;.
\end{equation}
The Lax pair matrices are again given by \eqref{LgeodLaxPairCoor}.
In the special case when ${h_{ab}}$ is nondegenerate, its very existence guarantees complete integrability of the geodesic motion, see the following subsection.

(c) The KY tensor ${\ph_{abc}}$ of rank 3 generates the Lax tensor
\begin{equation}\label{Lfp}
    L^a{}_{\!b} = \ph^a{}_{bn}\, p^n \;.
\end{equation}
This case has been discussed in papers \cite{Rosquist:1994,RosquistGoliath:1998, KarloviniRosquist:1999, BaleanuKarasu:1999, BaleanuBaskal:2000}.

(d) The KY tensor ${\ph_{ab}}$ of rank 2 generates the Lax tensor
\begin{equation}\label{LPhi}
    L^a{}_{\!b} = \Phi^a{}_{\!b} = \ph^{an} p_n p_b + p^a p^n \ph_{nb} \;.
\end{equation}
The invariant generated from this Lax tensor is a function of the observable\footnote{Here we used that ${p^2}$ is also the conserved quantity and we have canceled it out from ${\tr(L^2)}$.} $\!{p_a p_b k^{ab}}$, with ${k^a{}_b = \ph^{an}\ph_{bn}}$ being the Killing tensor of rank 2 associated with the KY tensor ${\ph_{ab}}$.

(e) Any rank two tensor constructed from the momentum ${p}$ and quantities $\kappa$ and $\mu$ (for various KY forms ${\ph}$ and CCKY forms ${h}$) by contractions and wedge operation. A simple interesting example is a ``partial square'' of the quantity $\kappa$, namely the Lax tensor
\begin{equation}\label{Lfpfp}
    L^a{}_{\!b} = \ph^{akl\dots m} p_m\, \ph_{bkl\dots n}  p^n\;.
\end{equation}
Trace of this Lax tensor gives the quadratic conserved observable ${p_a p_b k^{ab}}$, where the second rank Killing tensor ${k^{ab}}$ is associated with the KY form ${\ph}$:
\begin{equation}\label{KTKY2}
    k^a{}_b = \ph^{akl\dots} \, \ph_{bkl\dots}\;.
\end{equation}
Another interesting possibility is to take a `square' of tensor \eqref{PhiFdef},
\begin{equation}\label{LFsqr}
L^a{}_b=F^{akl\dots} F_{bkl\dots}\,.
\end{equation}
Depending on the rank of $\ph$ this will generate certain number of conserved quantities.

To summarize, the Lax tensors build from the KY and CCKY forms can be very fruitful. They can generate plenty of functionally independent invariants and they are thus very useful when investigating the conserved quantities.

\subsection*{Kerr-NUT-(A)dS spacetime}
A highly nontrivial example of the Lax tensor discusssed above can be found in the spacetime equipped with a nondegenerate CCKY tensor ${h_{ab}}$ of rank 2. It was proved in \cite{HouriEtal:2007, KrtousEtal:2008, HouriEtal:2008b} that the existence of such a principal CCKY tensor determines the form of the metric up to a set of metric functions of a single argument. These functions can be fixed by the Einstein equation and the resulting vacuum (with cosmological constant) metric describes a generally rotating black hole in arbitrary number of spacetime dimensions, also  called the Kerr-NUT-(A)dS spacetime \cite{MyersPerry:1986, GibbonsEtal:2004, ChenEtal:2006cqg}.

In this case, the principal CCKY tensor ${h_{ab}}$ generates the covariantly conserved Lax tensor ${F=(h\wedge p)\cdot p}\,$, cf.\ also \eqref{LF}, which in ${D=2n+\eps}$ $({\eps=0,1})$ dimensions generates ${n}$ independent constants of geodesic motion \cite{PageEtal:2007, KrtousEtal:2007jhep, KrtousEtal:2007prd, HouriEtal:2008a}. It is possible to choose these constants in such a way that they are quadratic in momentum and hence generated by rank 2 Killing tensors ${k_{(j)}^{ab}}$. Namely, it was shown in \cite{KrtousEtal:2007jhep} that for any parameter ${\beta}$ the following identity holds:
\begin{equation}\label{genKT}
    p^2\det\Bigl(I+\sqrt{\beta}\,p^{-2} F\Bigr) = \sum_{j=0}^{n} p_a p_b k_{(j)}^{ab}\, \beta^j\;.
\end{equation}
The left-hand side is a scalar expression, constructed just from the Lax tensor ${F^a{}_{\!b}}$ and ${p^2}$, and hence is conserved for any ${\beta}$. The coefficients in the ${\beta}$-expansion are thus also conserved and can be read from the right-hand side of \eqref{genKT}. They are quadratic in momentum with the ${k_{(j)}}$ being Killing tensors.\footnote{%
For ${j=0}$ the Killing tensor reduces to the metric, ${k_{(0)}^{ab}=g^{ab}}$. The ${n}$-th Killing tensor vanishes in even dimensions, whereas it is reducible to a square of one of the Killing vectors  in odd dimensions. For ${j=0,\dots,n-1}$ the Killing tensors are irreducible, giving thus ${n}$ quadratic constants of motion.}

Moreover, the discussed spacetime admits also ${n+\eps}$ explicit (Killing vector) symmetries which supply additional ${n+\eps}$ conserved quantities. All these ${D}$ conserved quantities are in involution and the system is completely integrable. The existence of the quadratic conserved quantities encoded in the Lax tensor is also a  starting point for showing that the Hamilton--Jacobi, Klein--Gordon, and Dirac equations separate in these spacetimes
\cite{FrolovEtal:2007, SergyeyevKrtous:2008, OotaYasui:2008, CarigliaEtal:2011a, CarigliaEtal:2011b}. (See also \cite{KunduriEtal:2006, OotaYasui:2010} for separability of
certain gravitational perturbations.)

\section{Motion of a charged particle}\label{sc:charged}

It was shown in \cite{FrolovKrtous:2011} that the motion of a charged particle in the special test electromagnetic field in the background of
Kerr-NUT spacetimes in all dimensions
is also completely integrable. Here we demonstrate that the conserved quantities quadratic in momentum can be found using the Lax tensor method.

The electromagnetic field under investigation is given by the vector potential ${A}$ proportional to the primary Killing vector ${\xi}$ of the Kerr-NUT geometry.\footnote{%
The condition of vanishing electric current for such an electromagnetic field requires the cosmological constant to be set equal zero, cf.~\cite{FrolovKrtous:2011}.}
The primary Killing vector ${\xi}$ is a divergence of the principal CCKY tensor ${h_{ab}}$, $\xi^a=\frac{1}{D-1}\nabla_ch^{ca}$, cf.\ Eq.~\eqref{CCKYdef}. We will write ${qA_a = e\xi_a}$, where ${q}$ is a charge of the particle and ${e}$ a constant combining both the charge and strength of the field. The motion of the charged particle is thus governed by the Hamiltonian:
\begin{equation}\label{HEM}
H=\frac{1}{2m}(p_a-e\xi_a) g^{ab}(p_b-e\xi_b)\;.
\end{equation}

The relation between the velocity and momentum can be read from the Hamiltonian flow \eqref{hamflowsplit}:
\begin{equation}\label{HamEq1EM}
u^a=\frac1m\,(p^a-e\xi^a)\;.
\end{equation}
The covariant derivative of the momenta is
\begin{equation}\label{HamEq2EM}
\frac{\nabla}{dt} p_a= \frac{e}{m} \bigl(\nabla_{\!a} \xi_n\bigr) (p^n-e\xi^n)\;.
\end{equation}
Substituting \eqref{HamEq1EM} and using the Killing vector condition ${\nabla_{\!a}\xi_n = -\nabla_{\!n}\xi_a}$ one gets ${\frac{\nabla}{dt} p_a}{=}{ -e u^n \nabla_{\!n} \xi_a}{=}{- e\frac{\nabla}{dt} \xi_a}$, i.e.,
\begin{equation}\label{p+xiCons}
\frac{\nabla}{dt} (p+e\xi)= 0\;.
\end{equation}

Now we can prove that the quantity
\begin{equation}
\mu={h\wedge(p+e\xi)}
\end{equation}
 is covariantly conserved. Indeed, using \eqref{p+xiCons}, \eqref{CCKYdef}, and \eqref{HamEq1EM}, we get
\begin{equation}\label{hpxiCons}
  \frac{\nabla}{dt} \mu = (u\wedge\xi)\wedge (p+e\xi) = 0\;.
\end{equation}
It follows  that
\begin{equation}\label{FdefEM}
  F = \mu \cdot(p+e\xi)
\end{equation}
is also covariantly conserved. Being the tensor of rank~2, this is the covariant Lax tensor for our system. It is constructed in a similar way as the Lax tensor from the previous subsection, only with substitution ${p\to p+e\xi}$. The scalars generated from ${F}$ can thus be read again from \eqref{genKT}. The quadratic constants of motion are
\begin{equation}\label{quadrEM}
    \tilde{K}_{(j)} = (p_a+e\xi_a)(p_b+e\xi_b)k_{j}^{ab}\;.
\end{equation}
These differ from the constants ${K_{(j)}}$ introduced in \cite{FrolovKrtous:2011}, but only by terms ${L_{(j)}}$ linear in momentum which are also conserved:
\begin{equation}\label{consEM}
\begin{split}\raisetag{3ex}
    K_{(j)} &= (p_a-e\xi_a)(p_b-e\xi_b)k_{j}^{ab} = \tilde{K}_{(j)} - 4 e\, L_{(j)}\;,\\
    L_{(j)} &= p_a l_{(j)}^a = p_a k_{(j)}^{an}\xi_n\;.
\end{split}
\end{equation}
The conservation of ${L_{(j)}}$ follows from the fact that ${l_{(j)}^a = k_{(j)}^{an}\xi_n}$ are Killing vectors \cite{KrtousEtal:2007jhep}. It was demonstrated in \cite{FrolovKrtous:2011} that the conserved quantities ${K_{(j)}}$ and ${L_{(j)}}$ are all in involution.

\section{Summary}\label{sc:summary}
The Lax pair formalism provides an elegant and effective description of special dynamical systems with enhanced symmetries. In particular, the existence of the Lax pair, defined by Eq.~\eqref{Lax}, enables one to generate constants of motion by simple algebraic operations, e.g., Eq.~\eqref{traces}.

In this paper we have provided an alternative, covariant formulation of the Lax formalism. This is based on the covariant (Clifford) Lax tensor, where the Lax equation is formulated as a covariant conservation of this tensor, Eqs.~\eqref{CovLaxEq} and \eqref{CovConsCl}. In both instances the existence of the Lax tensor enables one to generate constants of motion, which are determined as invariants constructed from the object, e.g., Eq.~\eqref{traces2}. We have further demonstrated that the ordinary Lax pair matrices follow from the covariant Lax formalism, the relation being given by Eqs. \eqref{LaxPairCoor} and \eqref{LaxPairCl}.

To illustrate the derived formulas, we have concentrated on the problem of particle motion in curved spacetime. In this case we were able to provide a number of examples of (Clifford) Lax tensors. In particular, we have concentrated on manifolds with enhanced symmetry, admitting hidden symmetries of Killing--Yano tensors, in which case the examples of Lax tensors are highly non-trivial. One of the Lax tensors discussed was proved responsible for complete integrability of geodesic motion in rotating black hole spacetimes in all dimensions just a few years ago. We have demonstrated for the first time that the conserved quantities for motion of a charged particle in the aligned test electromagnetic field on the same (vacuum) black hole background can be also generated using the Lax tensor \eqref{FdefEM}.

It remains an interesting open question whether any of the Lax tensors discussed here will find further physical applications in the future.

\vspace*{0.5cm}

\section*{Acknowledgments}
V.F. thanks the Natural Sciences and Engineering Research Council of Canada and the Killam Trust for the financial support.
P.K. was supported by Grants GA\v{C}R~202/09/0772 and GA\v{C}R~P203/12/0118.
D.K. and P.K. acknowledge hospitality at the University of Alberta where this work was partially done.
M.C. is partially funded by Fapemig under the project CEX APQ 2324-11.

\appendix

\section{Derivatives on the phase space with cotangent bundle structure}\label{apx:PhSpSpl}

In this Appendix we discuss the structure of the cotangent bundle phase space\footnote{%
We use capital Latin indices for the phase-space tensors and in this Appendix we write these indices explicitly. The material presented here partially follows and partially generalizes the Appendix of \cite{KrtousEtal:2007prd}.}
${P=\mathbf{T}^*M}$ in more details. We show that the covariant derivative on the configuration space ${M}$ induces a covariant splitting of the phase-space quantities into quantities related to the configuration space. It naturally replaces standard coordinate expressions in a coordinate independent way.

\subsection*{Derivatives along position and momentum directions}

First, we introduce covariant partial derivatives of a scalar observable along position and momentum directions. The derivative in momentum direction ${f_a}$ (i.e., changes along a curve ${p_a\to p_a+\eps f_a}$, ${x}$ fixed) is simple, since the space of momenta at fixed ${x}$ is linear. We define
\begin{equation}\label{apx:momder}
    f_a \frac{\partial F}{\partial p_a} = \frac{d}{d\eps} F(x,p+\eps f) \Big|_{\eps=0}\;.
\end{equation}
Thanks to ultralocality in ${f_a}$ we can tear off ${f_a}$ to obtain derivative operator ${\frac{\partial}{\partial p_a}}$ (with one contravariant configuration-space index) acting on scalar phase-space observables. Such a derivative operator also defines a mixed tensor ${\frac{\partial^A}{\partial p_a}\in \mathbf{T} P \otimes \mathbf{T}M}$. It is actually the tensor identifying the tangent space of the cotangent fibre ${\mathbf{T}(\mathbf{T}^*_x M)}$ with the cotangent fibre  ${\mathbf{T}^*_x M}$ itself.

The derivative along a position direction with momentum fixed is more involved. Moving from one position to another one changes the cotangent fibre and it has to be clarified what ``fixed momentum'' means. A natural solution is given in terms of a spacetime covariant derivative\footnote{It can be an arbitrary covariant derivative. Of course, in most cases it is useful to chose the metric derivative.} ${\nabla}$. The covariant derivative defines ``fixed momentum'' to be the parallel-transported momentum. Let ${x_\eps}$ be a spacetime curve in the ${u^a}$ direction and ${\bar p_\eps}$ be the parallelly transported momentum along this curve. Then we can write
\begin{equation}\label{apx:posder}
    u^a \frac{\nabla_{\!a} F}{\partial x}
       = \frac{d}{d\eps} F(x_\eps,{\bar p}_\eps) \Big|_{\eps=0}\;.
\end{equation}
Again, it defines the phase space derivative ${\frac{\nabla_{\!a}}{\partial x}}$ with one covariant configuration-space index and the mixed tensor ${\frac{\nabla^{\!A}_{\!a}}{\partial x}\in\mathbf{T}P\otimes\mathbf{T}^*M}$. The last quantity is the tensor which makes a horizontal lift of the configuration-space vector ${u^a}$ to the horizontal phase-space vector ${u^a \frac{\nabla^{\!A}_{\!a}}{\partial x}}$, see Fig.~\ref{fig:phspspl}.

The action of these derivatives on an observable of the form ${A(x,p)=\alpha^{ab\dots}(x)\,p_a p_b\dots}$ is
\begin{align}
    \frac{\nabla_{\!n} A}{\partial x} &=
      \nabla_{\!n}\alpha^{ab\dots}(x)\, p_a p_b \dots\;,
      \label{apx:posderact}\\
    \frac{\partial A}{\partial p_n} &=
      \alpha^{nb\dots}(x)\, p_b \dots + a^{an\dots}(x)\, p_a \dots + \dots\;.
      \label{apx:momderact}
\end{align}
The action on a general observable can be written explicitly using linearity, the Leibnitz product rule, and the chain rule.

The mixed tensors ${\frac{\nabla^{\!A}_{\!a}}{\partial x}}$ and ${\frac{\partial^A}{\partial p_a}}$ are covariant generalization of the phase-space coordinate vectors\footnote{%
The notation here is a bit mistreating the difference between covariant nature of the mixed tensors and coordinate-dependent nature of the coordinate tensors. For example, both indices in ${\frac{\nabla^{\!A}_{\!a}}{\partial x}}$ are tensor indices and could be understood as  abstract indices, independent of the chosen coordinates. Similarly for the phase-space index ${\scriptstyle A}$ in the coordinate vector ${\frac{\partial^A}{\partial x^n}}$. However, the index ${\scriptstyle n}$ here is not a tensor index, it just labels which coordinate tensor we are choosing. To make this distinction clear, one should distinguish the abstract and coordinate indices as, e.g., in \cite{PenroseRindler:book}. We decided not do so and let the reader distinguish tensorial and coordinate indices based on the context. In general, in expressions not involving explicitly chosen coordinates all indices are tensorial. If the coordinates are involved, the combinations ${x^a}$ and ${p_a}$ usually indicate the coordinate indices.}  ${\frac{\partial^A}{\partial x^a}}$ and ${\frac{\partial^A}{\partial p_a}}$ associated with the canonical coordinates ${(x^a,\, p_a)}$. Therefore, one can expect that natural symplectic quantities can be written using these covariant tensors. Namely, the inverse symplectic structure ${\Omega^{-1AB}}$ (such that ${\Omega_{AN} \Omega^{-1BN} = \delta^B_A}$), the Poisson brackets, and the Hamiltonian flow ${X_{\!\textstyle H}^A=\Omega^{-1AN}d_N H}$ are:
\begin{align}
  &\Omega^{-1\,AB}
    = \frac{\nabla^{\!A}_{\!n}}{\partial x}\,\frac{\partial^B}{\partial p_n}
    - \frac{\partial^A}{\partial p_n}\,\frac{\nabla^{\!B}_{\!n}}{\partial x}
    + p_n T^n_{\!kl} \frac{\partial^A}{\partial p_k} \frac{\partial^B}{\partial p_l}
    \;,\label{apx:InvSymplStrSpl}\\
  &\{F,G\}
    = \frac{\nabla_{\!n} F}{\partial x}\frac{\partial G}{\partial p_n}
    - \frac{\partial F}{\partial p_n}\frac{\nabla_{\!n} G}{\partial x}
    + p_n T^n_{\!kl} \frac{\partial F}{\partial p_k} \frac{\partial G}{\partial p_l}
    \;,\label{apx:PoisBrSpl}\\
  &X_{\!\textstyle H}^A
    = \frac{\partial H}{\partial p_n}\,\frac{\nabla^{\!A}_{\!n}}{\partial x}
    - \frac{\nabla_{\!n} H}{\partial x}\,\frac{\partial^A}{\partial p_n}
    - p_n\frac{\partial H}{\partial p_k} T^n_{\!kl} \frac{\partial^A}{\partial p_l}
    \;.\label{apx:hamflowsplit}
\end{align}
Here, ${T^n_{\!kl}}$ is the torsion of ${\nabla}$. For vanishing torsion ${T=0}$ the expressions resemble the standard coordinate formulae.

\subsection*{Dual quantities and splitting of the phase-space direction}

We can also construct dual quantities ${D_{\!A}^n\, x}$ and ${\nabla_{\!\!A}\, p_n}$  which corresponds to coordinate forms ${d_A x^a}$ and ${d_A p_a}$. We require the duality conditions
\begin{equation}\label{apx:dualitycond}
\begin{aligned}
    &\frac{\nabla^{\!N}_{\!a}}{\partial x}\; D^{b}_{\!N}x = \delta_a^b\;,
    &&\frac{\partial^{N}}{\partial p_a}\; \nabla_{\!\!N}p_b = \delta^a_b\;,\\
    &\frac{\nabla^{\!N}_{\!a}}{\partial x}\; \nabla_{\!\!N}p_b = 0\;,
    &&\frac{\partial^{N}}{\partial p_a}\; D^{b}_{\!N}x = 0\;.
\end{aligned}
\end{equation}
The completeness relation reads
\begin{equation}\label{apx:complrel}
    D_{\!A}^n x \; \frac{\nabla^{\!B}_{\!n}}{\partial x}
    + \nabla_{\!\!A} p_n \; \frac{\partial^{B}}{\partial p_n} = \delta_A^B\;.
\end{equation}
The symplectic structure ${\Omega_{AB}}$ and the symplectic potential ${\theta_A = - p_n d_A x^n }$ have a familiar form (except for the anomalous torsion term):
\begin{gather}
    \Omega_{AB} = D_{\!A}^n x \,\nabla_{\!\!B}p_n {-} \nabla_{\!\!A}p_n\, D_{\!B}^n x
      {-} p_n T^n_{\!kl}\, D^k_{\!A} x\, D^l_{\!B} x \label{apx:SymplStrSpl}\;,\\
    \theta_A = - p_n \, D^n_A x\label{apx:SymplPot}\;.
\end{gather}

\begin{figure}
\centerline{\includegraphics[width=320pt]{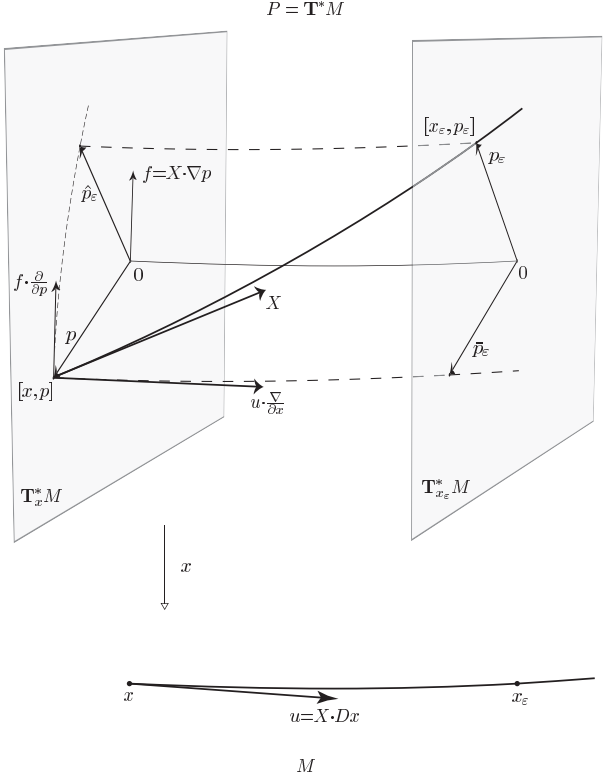}}
\vspace{-3ex}
\caption{\label{fig:phspspl}\textbf{Splitting of the phase-space direction ${X}$.}
In this figure we illustrate the splitting of the phase-space direction ${X}$ into its position and momentum parts.
Let ${[x_\eps,p_\eps]}$ be a curve starting at ${[x,p]}$ to which ${X}$ is tangent. Its corresponding position and momentum parts are ${x_\eps}$ and~${p_\eps}$. Dashed curve ${{\bar p}_\eps}$ is the parallel transport of the initial momentum ${p}$ along ${x_\eps}$. Short-dashed curve ${{\hat p}_\eps}$ laying in the cotangent fibre ${\mathbf{T}^*_x M}$ is obtained as the parallel transport of ${p_\eps}$ along ${x_\eps}$ back to the point ${x}$.
Various vectors defined in the text are tangent vectors to these curves. The phase-space vector ${X}$ is tangent to ${[x_\eps,p_\eps]}$, its position direction ${u}$ is tangent to ${x_\eps}$, the momentum part ${f}$ is the derivative of ${{\hat p}_\eps}$ understood as a cotangent vector. In other words, ${f}$ is the covariant derivative of ${p_\eps}$ along ${x_\eps}$, i.e., along the direction ${u}$. The formula \eqref{apx:Xsplit} splits ${X}$ into ${u\cdot\frac{\nabla}{\partial x}}$, which is tangent to~${{\bar p}_\eps}$, and into ${f\cdot\frac{\partial}{\partial p}}$, which is tangent to~${{\hat p}_\eps}$.}
\end{figure}

These dual mixed tensors define also the splitting of a phase space vector ${X^{\!A}}$ into configuration-space quantities ${u^a}$ and ${f_a}$ discussed in Sec.~\ref{sc:CovDerPhSp} near Eq.~\eqref{fieldexampl}. Such splitting is based on the observation that the covariant derivative splits the the tangent fibre ${\textbf{T}_{[x,p]}P}$ into horizontal and vertical subspaces.\footnote{%
The horizontal subspace of ${\textbf{T}_{[x,p]}P}$ gives the directions of parallel-transported momenta, the vertical subspace is tangent to the fibre ${\textbf{T}^*_x M}$.}
The configuration direction ${u^a}$ is the projection of ${X^{\!A}}$ onto spacetime (encoding thus the horizontal part) and the momentum direction ${f_a}$ is the vertical part of ${X^{\!A}}$. These two parts can be written using ${Dx}$ and ${\nabla p}$, see Fig.~\ref{fig:phspspl}.

Indeed, the tensor ${Dx}$ is just the differential of the projection map ${x: [x,p]\to x}$ from the phase space to the configuration space. Given a phase-space vector ${X^{\!A}}$, its shadow on the configuration space thus is
\begin{equation}\label{apx:Xu}
    u^a = X^{\!A} \; D_{\!A}^a x\;.
\end{equation}
On other hand, the quantity ${\nabla p}$ is the projector of a phase space vector ${X^A}$ on its vertical part ${f_a}$.
\begin{equation}\label{apx:Xf}
    f_a = X^{\!A} \; \nabla_{\!\!A} p_a\;.
\end{equation}

The completeness relation \eqref{apx:complrel} together with \eqref{apx:Xu} and \eqref{apx:Xf} gives
\begin{equation}\label{apx:Xsplit}
    X^{\!A} = u^n \frac{\nabla^{\!A}_{\!n}}{\partial x}
          + f_n \frac{\partial^A}{\partial p_n}\;,
\end{equation}
cf.\ Fig.~\ref{fig:phspspl}, and the derivative of scalar the observable ${F(x,p)}$ along the phase-space direction ${X^{\!A}}$ is thus
\begin{equation}\label{apx:PhSpDer}
    X^{\!A}\, d_A F
       = u^n \frac{\nabla_{\!n} F}{\partial x}
       + f_n \frac{\partial F}{\partial p_n}\;.
\end{equation}

\subsection*{The covariant derivative induced on the phase space}

The covariant derivative of the phase space fields with configuration space indices introduced in Sec.~\ref{sc:CovDerPhSp} can be understood as a generalization of the formula \eqref{apx:PhSpDer} to tensor fields. The derivative ${\frac{\partial}{\partial p_n}}$ acting in the momentum directions has the same definition \eqref{apx:momder}. The definition of the derivative ${\frac{\nabla_{\!a}}{\partial x}}$ in the configuration direction changes just employing the covariant derivative\footnote{\label{fnt:curvedep}%
Since ${A^{a\dots}_{b\dots}(x_\eps,p_\eps)}$ is just ${\eps}$-dependent, it can be understood as a tensor field along the configuration curve ${x_\eps}$ and just the standard covariant derivative in the configuration space is involved on the right-hand side of the definition \eqref{apx:covposder}.} in the definition~\eqref{apx:posder}
\begin{equation}\label{apx:covposder}
    u^n \frac{\nabla_{\!n} A^{a\dots}_{b\dots}}{\partial x}
       = \frac{\nabla}{d\eps} A^{a\dots}_{b\dots}(x_\eps,{\bar p}_\eps) \Big|_{\eps=0}\;.
\end{equation}

The action of the derivatives ${\frac{\nabla_{\!a}}{\partial x}}$ and ${\frac{\partial}{\partial p_n}}$ on tensor field ${A^{a\dots}_{b\dots}(x,p) = \alpha^{a\dots kl\dots}_{b\dots}\,p_kp_l\dots}$ is analogous to rules \eqref{apx:posderact} and \eqref{apx:momderact}, just with additional indices involved.

The covariant derivative along a general phase-space direction ${X^A}$ (split as in \eqref{apx:Xsplit}) acting on a field ${A^{a\dots}_{b\dots}(x,p)}$ is then given by generalization of \eqref{apx:PhSpDer}:
\begin{equation}\label{apx:DerSpl}
   \frac{\nabla}{d\eps} A^{a\dots}_{b\dots}
      \equiv \nabla_{\!\textstyle X}\,A^{a\dots}_{b\dots}
      = u^n \frac{\nabla_{\!n} }{\partial x} A^{a\dots}_{b\dots}
       + f_n \frac{\partial }{\partial p_n} A^{a\dots}_{b\dots}\;.
\end{equation}
This we have already mentioned in Eq.~\eqref{DerSpl}. In particular, the derivatives of the pure configurations field ${A^{a\dots}_{b\dots}(x)}$ and of the momentum field ${p_a}$ are given by rules (i) and (ii) in Sec.~\ref{sc:CovDerPhSp}.

Since the dependence on ${X^{\!A}}$ is ultralocal, it is possible to define also the covariant differential ${\nabla_{\!\!N}A^{a\dots}_{b\dots}}$, namely
\begin{equation}\label{apx:CovPhSpDif}
    \nabla_{\!\!N}A^{a\dots}_{b\dots}
      = D_{\!N}^n x \,\frac{\nabla_{\!n}}{\partial x} A^{a\dots}_{b\dots}
      + \nabla_{\!\!N} p_n \,\frac{\partial}{\partial p_n} A^{a\dots}_{b\dots}\;.
\end{equation}

\subsection*{Relation to canonical coordinates}

In the definitions above we have not used any specific choice of the coordinates. However, the introduced formalism can be easily accommodated to such a choice. If we choose configuration-space coordinates ${x^a}$, one can define the ``coordinate derivative'' ${\partial}$ by conditions
\begin{equation}\label{apx:coorder}
    \partial d x^b =0\;, \quad \partial \frac{\partial}{\partial x^b} = 0\;.
\end{equation}
It is a torsion-free covariant derivative (of course, depending on the choice of coordinates). The difference tensor between ${\nabla}$ and ${\partial}$ is given by the connection coefficients~${\Gamma^n_{\!ab}}$.

We can use the coordinate derivative ${\partial}$ instead of ${\nabla}$ in all expressions above. It leads to standard coordinate expressions: the equation \eqref{apx:SymplStrSpl} reduces to \eqref{SymplStrCoor}, the expression \eqref{apx:PoisBrSpl} to \eqref{PoisBrCoor}, etc. In the coordinate case we also use the more common notation ${\frac{\partial}{\partial x^a}}$ instead of ${\frac{\partial_a}{\partial x}}$. The induced coordinate derivative on  phase space along a phase-space direction ${X^{\!A}}$ is denoted just by dot
\begin{equation}\label{apx:dotder}
    {\dot A}^{a\dots}_{b\dots} = \frac{\partial}{\partial\eps} A^{a\dots}_{b\dots}
     = \partial_{\textstyle X} A^{a\dots}_{b\dots}\;.
\end{equation}

Splitting of the phase-space direction ${X^{\!A}}$ depends on the choice of the covariant derivative. For the coordinate derivative this splitting gives:
\begin{equation}\label{apx:xiuudot}
    u^a = \xi^A \; D_{\!A}^a x\;, \qquad
    \dot p_a = \xi^A \; \partial_{\!A} p_a\;.
\end{equation}
Clearly, ${\dot p_a}$ are just derivatives of components of ${p}$ along ${X^{\!A}}$ -- which justifies the dot notation.

Let ${[x_\eps,p_\eps]}$ be a phase-space curve with tangent vector ${X}$. Then, ${f_a = X^{\!A} \nabla_{\!A}p_a}$ can be understood as the standard configuration-space covariant derivative of ${p_\eps}$ along the spacetime curve ${x_\eps}$, cf.\ Fig.~\ref{fig:phspspl}. Similarly, ${{\dot p}_a=X^{\!A} \partial_{\!A}p_a}$ is the coordinate derivative of ${p_\eps}$ along ${x_\eps}$. We can thus use the ordinary relation between covariant and coordinate derivatives to obtain
\begin{equation}\label{apx:fpdotrel}
   f_a = {\dot p}_a - u^k \,\Gamma_{\!ka}^{\,l} \,p_l\;,
\end{equation}
cf.~\eqref{finCoor}. Here, the velocity ${u^k}$ is tangent to the curve~${x_\eps}$.

Tearing off the phase-space vector ${X^A}$, we obtain
\begin{equation}\label{apx:nablaptransf}
    \nabla_{\!A} p_n = \partial_{\!A} p_n - D_{\!A}^k x\,\Gamma_{\!kn}^{\,l} \,p_l\;.
\end{equation}
The duality relations \eqref{apx:dualitycond} imply
\begin{equation}\label{apx:nabladxtransf}
    \frac{\nabla^{\!A}_{\!n}}{\partial x}
    = \frac{\partial^A}{\partial x^n}
    + p_k \,\Gamma_{\!nl}^k \,\frac{\partial^{\!A}}{\partial p_l}\;.
\end{equation}
The action on a scalar observable reads
\begin{equation}\label{apx:posdertrans}
    \frac{\nabla_{\!n} F}{\partial x}
    = \frac{\partial F}{\partial x^n}
    + p_k \,\Gamma_{\!nl}^k \,\frac{\partial F}{\partial p_l}\;.
\end{equation}
The generalization to the action on tensor fields adds only standard terms for each tensor index:
\begin{equation}\label{apx:poscovdertrans}
\begin{split}
    \frac{\nabla_{\!n}}{\partial x} A^{a\dots}_{b\dots}
    &= \frac{\partial  A^{a\dots}_{b\dots}}{\partial x^n}
    + p_k \,\Gamma_{\!nl}^k \,\frac{\partial A^{a\dots}_{b\dots}}{\partial p_l} \\
    &\quad
    + \Gamma_{\!nk}^a  \,A^{k\dots}_{b\dots} + \dots
    - \Gamma_{\!nb}^k  \,A^{a\dots}_{k\dots} - \dots
    \;.
\end{split}
\end{equation}

If we substitute \eqref{apx:fpdotrel} and \eqref{apx:poscovdertrans} into \eqref{apx:DerSpl}, we obtain the coordinate expression for the covariant derivative on the phase space
\begin{equation}\label{apx:CovPhSpDerTrans}
    \frac{\nabla}{\partial\eps} A^{a\dots}_{b\dots}
    = {\dot A}^{a\dots}_{b\dots}
       + u^n \Gamma_{\!nk}^a  \,A^{k\dots}_{b\dots} + \dots
       - u^n \Gamma_{\!nb}^k  \,A^{a\dots}_{k\dots} - \dots
\end{equation}
Here, the coordinate derivative ${{\dot A}^{a\dots}_{b\dots}}$ splits as
\begin{equation}\label{apx:CoorCovPhSpDer}
    {\dot A}^{a\dots}_{b\dots}
      = u^n \,\frac{\partial  A^{a\dots}_{b\dots}}{\partial x^n}
      + {\dot p}_n \,\frac{\partial A^{a\dots}_{b\dots}}{\partial p_n}\;.
\end{equation}

\vspace*{4ex}



%

\end{document}